\documentclass[lettersize,journal]{IEEEtran}
\usepackage{amsmath,amsfonts}
\usepackage{booktabs}
\usepackage{multirow}
\usepackage{array}
\usepackage[caption=false,font=normalsize,labelfont=normalfont,textfont=normalfont]{subfig}
\usepackage{textcomp}
\usepackage{stfloats}
\usepackage{url}
\usepackage{verbatim}
\usepackage{graphicx}
\usepackage{cite}
\usepackage{amssymb}
\usepackage[ruled,vlined]{algorithm2e}
\usepackage[table]{xcolor}
\usepackage{xcolor}
\usepackage{amsthm}

\newtheoremstyle{boldplain}
  {\topsep}
  {\topsep}
  {\normalfont}
  {}
  {\bfseries}
  {.}
  { }
  {}

\theoremstyle{boldplain}
\newtheorem{theorem}{Theorem}

\newtheorem{definition}{Definition}

\SetNlSty{\bfseries}{}{:\,}
\SetAlgoNlRelativeSize{0}
\SetNlSkip{2em}

\SetKwInput{KwInit}{Initialization}
\SetKwInput{KwOut}{Output}
\SetKw{KwEach}{Each}
\DontPrintSemicolon
\SetInd{0.5em}{1.0em}
\SetAlFnt{\small}
\SetKw{KwTo}{to}
\SetKwRepeat{Repeat}{repeat}{until}
\SetAlgoSkip{smallskip}

\begin{document}
\title{Mechanism and Communication Co-Design for Differentially Private Energy Sharing}

\author{Yingshuo Gu,
        Xi Weng,
        Yue Chen,~\IEEEmembership{Senior Member,~IEEE}
\thanks{Y. Gu and Y. Chen are with the Department of Mechanical and Automation Engineering, The Chinese University of Hong Kong, Hong Kong, China (e-mail: \{ysgu, yuechen\}@mae.cuhk.edu.hk).}
\thanks{X. Weng is with the Guanghua School of Management, Peking University, China (email: wengxi125@gsm.pku.edu.cn)}}

\markboth{Journal of \LaTeX\ Class Files,~Vol.~14, No.~8, August~2021}%
{Shell \MakeLowercase{\textit{et al.}}: A Sample Article Using IEEEtran.cls for IEEE Journals}


\maketitle

\begin{abstract}
Integrating distributed energy resources (DERs) is a critical step toward addressing the global climate crisis. This transformation has driven the transition from traditional consumers to prosumers and given rise to new energy sharing business models. Existing works have extensively studied prosumer energy sharing mechanisms, yet little attention has been paid to privacy protection, particularly when communication constraints are taken into account. In this paper, we study an energy sharing mechanism where information is exchanged over wireless channels via over-the-air (OTA) multiple-input multiple-output (MIMO) aggregation to exploit spectral efficiency for scalable prosumer coordination. To characterize the privacy leakage risk during data transmission process, we introduce an adversarial attack model and demonstrate that, under certain conditions, the platform can extract and recover prosumers' private parameters from the base station observations. To safeguard the energy sharing mechanism against such attacks, we propose a differentially private equilibrium-seeking algorithm, analyze the achievable privacy level, and establish convergence guarantees that quantify the impact of privacy on the convergence accuracy. Numerical experiments demonstrate that our approach effectively protects prosumers' privacy while converging to near-optimal solutions.
\end{abstract}


\section{Introduction}
\IEEEPARstart{T}{he} integration of distributed energy resources (DERs), such as rooftop solar panels, battery storage systems, and electric vehicles, is fundamentally transforming the landscape of modern power systems~\cite{basak2012literature}. This transformation has given rise to the concept of \emph{prosumers}---energy consumers who can also produce and store electricity locally. Unlike traditional consumers, prosumers actively participate in energy markets by trading their surplus generation or purchasing energy when needed. Energy sharing among prosumers enables efficient utilization of locally generated renewable energy, reduces reliance on centralized power plants, and promotes sustainability.

Multiple prosumers participating in energy sharing form a game, where each prosumer, acting as a strategic player, makes local decisions (production, consumption, and trading) to maximize individual payoff, and market-clearing prices coordinate these decentralized decisions toward equilibrium. Existing game-theoretic approaches typically assume perfect and instantaneous information exchange between prosumers/platform, under which convergence to the energy sharing game equilibrium can be guaranteed. In practice, however, this information exchange relies on wireless communication infrastructure that is subject to channel impairments, bandwidth limitations, and resource constraints, which cannot be ignored. In this paper, we consider over-the-air (OTA) computation with multiple-input multiple-output (MIMO) antenna arrays for energy sharing. OTA computation exploits the superposition property of wireless channels. It is a promising communication scheme that has been widely adopted, compared to conventional orthogonal multiple-access schemes, such as time division multiple access (TDMA) and frequency division multiple access (FDMA), that may incur significant bandwidth consumption and scale poorly as the number of prosumers grows~\cite{goldsmith2005wireless}.


During this wireless information exchange, a major concern is the potential privacy leakage of prosumers: Attackers may infer sensitive prosumer information from the nonsensitive data exchanged. To address this challenge, various approaches have been developed.
For example, cryptographic approaches such as homomorphic encryption and secure multi-party computation provide strong theoretical privacy guarantees~\cite{gentry2009fully,yao1982protocols}, but typically incur substantial computational and communication overhead and require complex multi-round interactions that are impractical for resource-constrained prosumer devices. In contrast, differential privacy (DP) offers provable privacy guarantees by injecting calibrated noise into transmitted information with relatively low computational cost~\cite{dwork2006calibrating}, which is the focus of this paper. However, most existing DP-based approaches overlook the inherent noise in wireless channels and assume ideal communication. This leads to injecting more artificial noise than necessary to meet privacy requirements, which compromises convergence accuracy. 

To fill the research gaps arising from the limited consideration of communication paradigms in energy sharing, this paper proposes a mechanism and communication co-design approach for ensuring differential privacy in energy sharing. Related research work is discussed below.
\subsection{Related Work}

\subsubsection{Energy Sharing Mechanism Design}
Energy sharing among prosumers with DERs has emerged as a promising paradigm to facilitate renewable integration and system decarbonization~\cite{tushar2020peer}. Game-theoretic approaches have been widely adopted for energy sharing mechanism design, including cooperative game-based approaches and non-cooperative game-based approaches \cite{chen2022review}. Cooperative game-based approaches focus on coalition formation and fair revenue allocation in local energy communities~\cite{malik2022priority,tushar2019motivational}. Non-cooperative game-based approaches include Stackelberg games and generalized Nash games. In a Stackelberg game, a platform or aggregator acts as the leader to set prices, while prosumers respond as followers to maximize individual payoffs~\cite{el2017managing,liu2017energy,wang2024game}. Generalized Nash games capture coupled constraints among prosumers, and distributed algorithms are developed to seek the generalized Nash equilibrium (GNE)~\cite{chen2019energy,chen2020approaching,wang2025online}. These works have established rigorous theoretical foundation for energy sharing. However, they all assume perfect communication and \textit{overlook the constraints imposed by realistic wireless communication channels}.


\subsubsection{Wireless Communication in Distributed Systems}
The iterative information exchange required by energy sharing games relies on communication infrastructure. Recent works have incorporated channel characteristics into distributed system design. Reference~\cite{ullah2021distributed} models communication network topology in energy trading, showing that network structure impacts consensus convergence. A communication reliability-aware energy sharing strategy is developed in~\cite{chen2021communication}, demonstrating that poor channel quality can degrade efficiency. OTA computation has emerged as a promising technique for low-latency aggregation by exploiting wireless superposition~\cite{mitsiou2023accelerating,lin2022distributed}. OTA MIMO frameworks are developed for IoT networks with optimized beamforming design~\cite{wen2019reduced}, and OTA-based federated learning is studied over device-to-device networks~\cite{xing2021federated}. While these studies demonstrate the importance of communication-aware design, their direct application to energy sharing remains limited, as energy sharing involves strategic interactions among self-interested prosumers and requires iterative equilibrium seeking rather than standard distributed optimization. Moreover, \textit{none of them address the privacy risks inherent in wireless information exchange}.

\subsubsection{Privacy Protection}
Differential privacy has emerged as a rigorous framework for quantifying and controlling privacy leakage in distributed systems~\cite{wei2021user}. DP-based distributed optimization algorithms are developed with gradient tracking, revealing the privacy-convergence dilemma~\cite{huang2024differential}. In energy trading, DP mechanisms are applied by injecting calibrated noise into exchanged information~\cite{cao2024differentially,liu2024differentially,wei2024peer}. Reference~\cite{wang2024differentially} develops differentially private Nash equilibrium seeking algorithms for network games. These works establish DP as a powerful tool for privacy protection, while highlighting the inherent tradeoff between privacy and convergence performance. However, they typically assume \textit{ideal communication}. To account for communication constraints, DP analysis of over-the-air federated learning in~\cite{park2023differential} quantifies the required artificial noise for privacy preservation, and the privacy-learning tradeoff over MIMO fading channels is characterized in~\cite{liu2024ota}. However, these two works cannot be directly applied to energy sharing, which has more complicated strategic interactions. The joint consideration of communication constraints and privacy protection in energy sharing remains largely unexplored.


\subsection{Contributions and Organization}
To fill the research gaps above, in this paper, we study a wireless energy sharing system where prosumers iteratively submit bids over OTA MIMO channels to a platform that computes and broadcasts market-clearing prices. Our main contributions are two-fold:

1) \emph{Mechanism-Communication Co-Design Framework}: We develop a novel framework for jointly designing market mechanism and communication schemes for energy sharing considering the privacy-accuracy tradeoff. Unlike conventional approaches that assume the platform receives exact bidding values from prosumers, our framework models the signal encoding, transmission, and decoding processes explicitly. The impact of inherent communication channel noise and externally injected artificial noise on the received bids is also characterized. Furthermore, we introduce an honest-but-curious adversarial model at the base station, which exploits MIMO spatial resolution to infer prosumers' private parameters from OTA observations. This reveals the necessity of privacy protection in wireless energy sharing systems.
  
2) \emph{Differentially Private Equilibrium-Seeking Algorithm}: We design a distributed algorithm where prosumers transmit bids via OTA MIMO aggregation with calibrated artificial noise to achieve $(\epsilon,\delta)$-differential privacy. We prove convergence of the algorithm to a Nash equilibrium under noisy channels, and analytically characterize the privacy-convergence tradeoff. 



The remainder of this paper is organized as follows. Section \ref{sec-II} introduces the energy sharing game. Section \ref{sec:OTA} presents the wireless communication model based on OTA MIMO transmission and the adversarial attack model to reveal privacy risks. Section \ref{sec-IV} develops a differentially private equilibrium-seeking algorithm and analyzes its properties. 
Section \ref{sec-V} presents the case studies. Section \ref{sec-VI} concludes the paper.

\section{Energy Sharing Game}
\label{sec-II}

We consider an energy sharing system where a platform coordinates $I$ prosumers indexed by $i\in\mathcal{I}=\{1,2,\ldots,I\}$. Each prosumer $i$ self-produces power $p_i$ at cost $f_i(p_i)$ and consumes $d_i$ to obtain utility $u_i(d_i)$. We assume that function $f_i(\cdot)$ is strictly convex, $u_i(\cdot)$ is strictly concave and both of them are twice differentiable.

For each prosumer $i\in\mathcal{I}$, let $q_i=d_i-p_i$ denote the traded energy, positive when buying and negative when selling. We adopt the intercept-function bidding scheme: prosumer $i$ submits a scalar bid $b_i$, which represents the prosumer's willingness to buy and the platform posts a uniform clearing price $\lambda$. Their relation is modeled by the generalized linear demand function\cite{hobbs2000strategic}:
\begin{equation}
    q_i = -a\lambda + b_i, 
\end{equation}
where $a>0$ is the market sensitivity. Market clearing requires $\sum_{i\in\mathcal{I}}q_i=0$, yielding the clearing price
\begin{equation}
    \lambda = \frac{1}{aI}\sum_{i\in\mathcal{I}}b_i.
    \label{equ:price}
\end{equation}

Based on the above setting, each prosumer aims to maximize its net payoff, defined as the utility from consumption minus production cost and the trading payment at the clearing price $\lambda$, subject to energy balance. Specifically, prosumer $i$ solves
\begin{subequations}
\label{equ:game}
\begin{align}
    \max_{p_i,d_i,b_i}\quad &U_i(p_i,d_i,b_i) \notag\\
    &: = u_i(d_i) - f_i(p_i)-\lambda(b)(-a\lambda(b)+b_i), \label{equ:3a}\\
    \text{s.t.}\quad &p_i - a\,\lambda(b) + b_i = d_i,\label{equ:3b}\\
    &\lambda(b) = \frac{\sum_{j\in\mathcal{I}}b_j}{aI}.\label{equ:3c}
\end{align}
\end{subequations}

\begin{figure*}[!t]
\centering
\includegraphics[width=\textwidth]{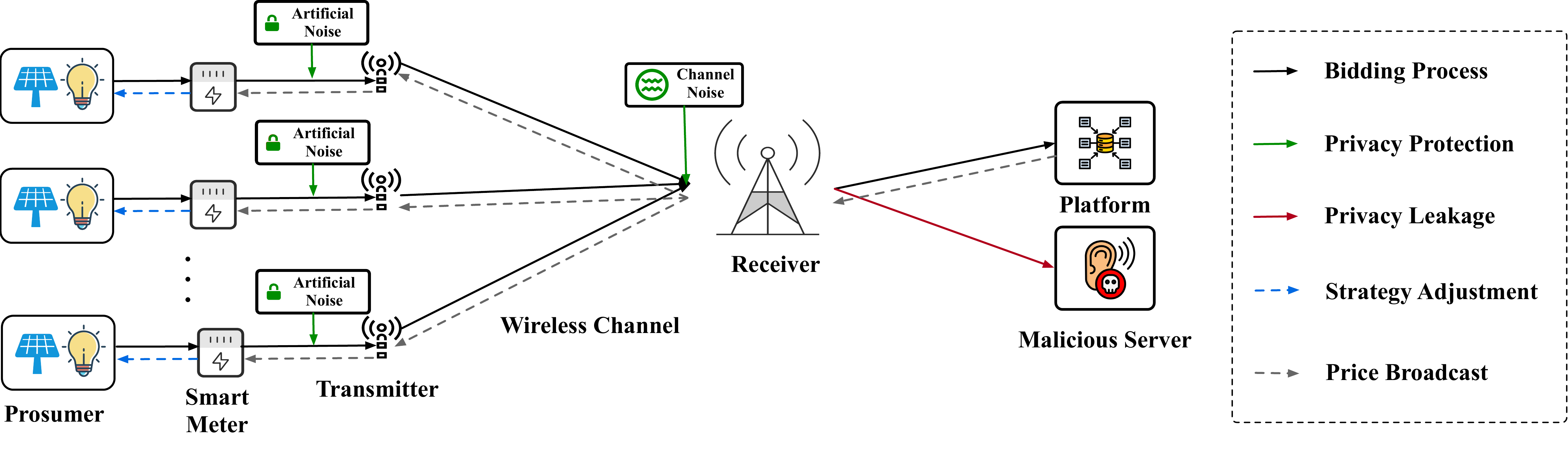}
\caption{System model of the OTA-based energy sharing framework, where prosumers submit bids over wireless MIMO channels to a platform that computes and broadcasts market-clearing prices.}
\label{fig:system_model}
\end{figure*}

Under this setting, all prosumers under energy sharing form a game, denoted by $\mathcal{G}=(\mathcal{I},\mathcal{S},U)$, where $\mathcal{I}:=\{1,\ldots,I\}$ is the set of prosumers, $\mathcal{S}:=\{\mathcal{S}_1,\ldots,\mathcal{S}_I\}$ is the collection of action sets, and $U:=\{U_1,\ldots,U_I\}$ is the collection of payoff functions. The action of player $i$ is $s_i:=(p_i,d_i,b_i)$, and the feasible action set $\mathcal{S}_i$ is defined by~\eqref{equ:3b}--\eqref{equ:3c}, influenced by the other players' bids $b_{-i}:=\{b_j\}_{j\neq i}$. Since the clearing price $\lambda(b)$ in both the objective \eqref{equ:3a} and the constraint \eqref{equ:3c} couples all prosumers' bids, the game $\mathcal{G}$ is a generalized Nash game, whose equilibrium is defined as follows.

\begin{definition}[Generalized Nash Equilibrium]
\label{def:GNE}
A strategy profile $(p^\star,d^\star,b^\star)\in\mathcal{S}$ is a GNE of the energy sharing game $\mathcal{G}$, if for all $i\in\mathcal{I}$:
\begin{align*}
    (p^\star_i,d^\star_i,b^\star_i) \in &\arg\max\; U_i\bigl(p_i,d_i,b_i,p^\star_{-i},d^\star_{-i},b^\star_{-i}\bigr),\\
    &\text{s.t. } \eqref{equ:3b}\text{--}\eqref{equ:3c}.
\end{align*}
\end{definition}

It has been shown in~\cite[Proposition~1]{chen2020approaching} that a GNE of $\mathcal{G}=(\mathcal{I},\mathcal{S},U)$ exists. Moreover, for any GNE $(p^\star,d^\star,b^\star)$ with clearing price $\lambda^\star$, the pair $(p^\star,d^\star)$ is the unique optimal solution to
\begin{subequations}
\label{equ:GNE_char}
\begin{align}
    \min_{p_i,d_i,\,\forall i\in\mathcal{I}}\; &\sum_{i=1}^{I}\bigl[f_i(p_i) - u_i(d_i)\bigr] + \frac{\sum_{i=1}^{I}(d_i-p_i)^2}{2a(I-1)}, \label{equ:GNE_obj}\\
    \textnormal{s.t.}\quad &\sum_{i=1}^{I} p_i = \sum_{i=1}^{I} d_i \,:\, \zeta, \label{equ:GNE_balance}
\end{align}
\end{subequations}
and the GNE price satisfies $\lambda^\star = \zeta^*$, where $\zeta^*$ is the value of dual variable of the power balance constraint~\eqref{equ:GNE_balance} at optimum. Consequently, the production and demand $(p_i^\star,d_i^\star)$ of each prosumer $i$ maximize the surrogate payoff
\begin{equation}
\label{equ:surrogate}
    U_i(p_i,d_i;\lambda) := u_i(d_i) - f_i(p_i) - \lambda(d_i\!-\!p_i) - \frac{(d_i-p_i)^2}{2a(I-1)}
\end{equation}
for $\lambda=\lambda^\star$, where the quadratic term accounts for each prosumer's price impact on the clearing price~\cite{chen2020approaching}.

In practice, the game is solved iteratively: the platform broadcasts a clearing price $\lambda$, each prosumer solves the local problem
\begin{equation}
\label{equ:local}
    \max_{p_i,d_i}\; U_i(p_i,d_i;\lambda),
\end{equation}
and the bid $b_i$ is determined by the energy balance~\eqref{equ:3b}. The platform then aggregates all bids to update the price via~\eqref{equ:price}.

\section{Communication-Aware Implementation}
\label{sec:OTA}
For the energy sharing game introduced in Section \ref{sec-II}, each prosumer $i$ needs to submit their bid $b_i$ to the platform. In this section, we model the underlying communication processes. We consider the wireless network depicted in Fig.~\ref{fig:system_model}, where the platform is implemented as an $N_r$-antenna base station (BS) and each prosumer is equipped with a single antenna. In the energy sharing game, there exist iterative exchanges of information between prosumers and the platform. In practice, these exchanges are carried over wireless channels: prosumers send their bids to the platform via the uplink channels, and the platform broadcasts the clearing price to all prosumers via the downlink channels. Since the downlink transmission is typically supported by sufficient power at the BS, we assume an error-free downlink broadcast, an assumption commonly used~\cite{liu2024differentially}. Therefore, in the following, we focus on the uplink transmission of bids.

\begin{figure}[t]
\centering
\includegraphics[width=\linewidth]{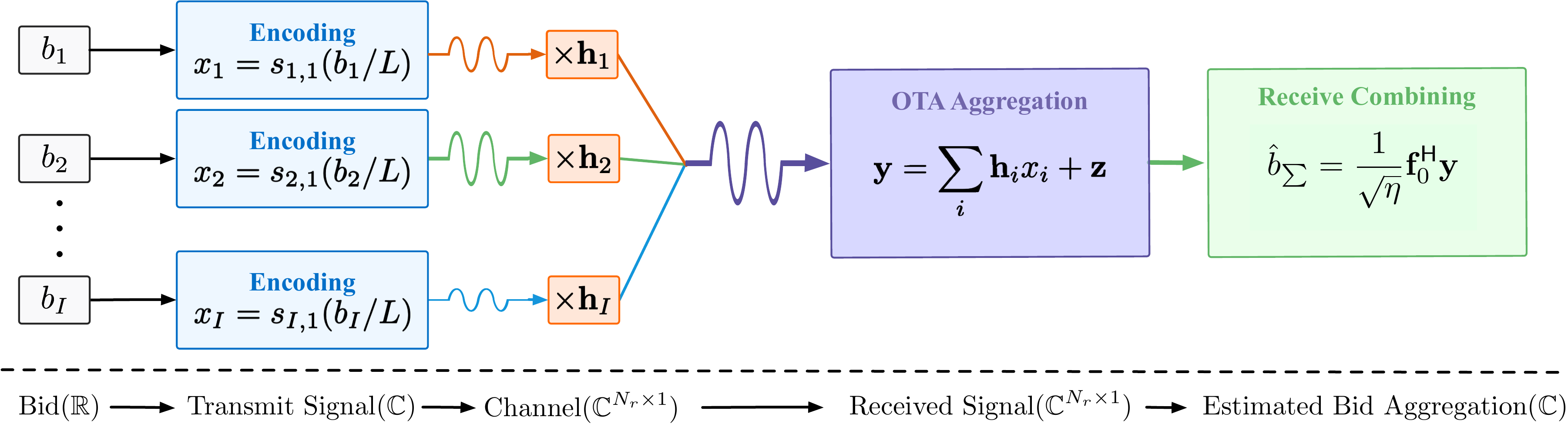}
\caption{Communication flow of the OTA-based bid aggregation.}
\label{fig:comm_flow}
\end{figure}

The objective of the uplink stage is to aggregate all prosumers' bids and compute the clearing price according to~\eqref{equ:price}. By exploiting superposition on the uplink multiple-access channel, OTA transmission naturally realizes this aggregation. The OTA-based uplink transmission procedure is illustrated in Fig.~\ref{fig:comm_flow}, with details provided as follows.

\subsection{Bid Encoding and Scaling}
Let $b_i^k$ denote the bid of prosumer $i$ in the $k$-th iteration. Assume a unified upper bound $L$ for the bids:
\begin{equation}
\label{eq:7}
    \lvert b_i^k\rvert\leq L,\forall i,k.
\end{equation}
Under this assumption, each prosumer $i\in\mathcal{I}$ encodes its bid $b_i^k$ in the $k$-th iteration into a complex baseband symbol $x_i^k$ under a transmit power constraint. Specifically, the transmitting signal of the $i$-th prosumer is set to
\begin{equation}
\label{equ:transmit}
   x_i^k = s_{i,1}\frac{b_i^k}{L},\quad\forall i,k,
\end{equation}
where $s_{i,1}\in\mathbb{C}$ is the transmit scalar. Each prosumer is subject to a transmit power constraint:
\begin{equation}
\label{equ:power_constraint_orig}
    |x_i^k|^2 = |s_{i,1}|^2\frac{|b_i^k|^2}{L^2} \leq P,\quad\forall i,k,
\end{equation}
where $P$ denotes the maximum transmit power budget. Since $|b_i^k/L|\leq 1$, constraint \eqref{equ:power_constraint_orig} is guaranteed by
\begin{equation}
    |s_{i,1}|^2\leq P,\quad\forall i.
\end{equation}

\subsection{Uplink Synchronization and OTA Aggregation}
Then we leverage the superposition property of wireless multiple-access channels to enable efficient OTA aggregation of the transmitted signals. Specifically, the BS is equipped with $N_r$ receive antennas, forming a MIMO uplink channel. In each iteration $k$, all prosumers simultaneously transmit their encoded bid signals $\{x_i^k\}_{i\in\mathcal{I}}$ using the same radio resource to ensure proper symbol synchronization. This property aligns well with the market clearing rule that requires computing the aggregate $\sum_i b_i^k$, allowing the BS to directly obtain the needed information without individually decoding each bid. Compared with conventional sequential reporting, this simultaneous transmission not only reduces communication latency but also inherently obfuscates individual bids, which offers an additional layer of privacy protection. The corresponding received signal $\mathbf{y}^k\in\mathbb{C}^{N_r\times 1}$ is given by
\begin{equation}
\label{equ:receive signal}
   \mathbf{y}^k = \sum_{i\in\mathcal{I}}\mathbf{h}_ix_i^k+ \mathbf{z}^k,
\end{equation}
where $\mathbf{h}_i\in\mathbb{C}^{N_r\times 1}$ is the uplink channel coefficient from prosumer $i$ to the platform. We assume that channels remain constant during the algorithm execution and that perfect channel state information is available at the BS. $\mathbf{H}\triangleq[\mathbf{h}_1,\ldots,\mathbf{h}_I]\in\mathbb{C}^{N_r\times I}$ denotes the channel matrix. The term $\mathbf{z}^k\in\mathbb{C}^{N_r\times1}$ models additive white complex Gaussian noise on the wireless link, with $\mathbf{z}^k\sim\mathcal{CN}(\mathbf{0},\sigma_z^2\mathbf{I})$, where $\sigma_{z}^2$ is the noise variance.

\subsection{Receive Combining and Normalization}
Given $\mathbf{y}^k$ in \eqref{equ:receive signal}, the platform forms a scalar estimate of the aggregate bid by applying a unit-norm combining vector $\mathbf{f}_0\in\mathbb{C}^{N_r\times1}$ (${\|\mathbf{f}_0\|_2^2}=1$):
\begin{equation}
\begin{aligned}
     \hat{b}_{\Sigma}^k&= \frac{1}{\sqrt\eta}\mathbf{f}_0^{\mathsf{H}}\mathbf{y}^k \\
     & = \sum_{i\in\mathcal{I}}\frac{\mathbf{f}_0^{\mathsf{H}}\mathbf{h}_is_{i,1}}{\sqrt{\eta}L} b_i^k + \frac{\mathbf{f}_0^{\mathsf{H}}\mathbf{z}^k}{\sqrt{\eta}},
\end{aligned}
\end{equation}
where $\eta>0$ normalizes the combining vector. The clearing price is then calculated as
\begin{equation}
\label{equ:estimate price}
\begin{aligned}
    \hat{\lambda}^k &= \frac{1}{aI}\hat{b}_{\Sigma}^k \\
    & = \sum_{i\in\mathcal{I}}\frac{\mathbf{f}_0^{\mathsf{H}}\mathbf{h}_is_{i,1}}{aI\sqrt{\eta}L} b_i^k + \frac{\mathbf{f}_0^{\mathsf{H}}\mathbf{z}^k}{aI\sqrt{\eta}},
\end{aligned}
\end{equation}
and broadcast back to all prosumers for decision updates.

\subsection{Potential Privacy Leakage at the Base Station}
\label{subsec:attack_model}
In the considered energy-sharing system, each prosumer holds private parameters that reflect its individual demand preferences. Specifically, the net demand $d_i - p_i$, representing the difference between local consumption and generation, is sensitive information that prosumers are not intended to disclose to the BS. 
Although OTA transmission aggregates individual bids into a single superposition, the MIMO array at the BS provides spatially diverse observations that can be exploited to separate individual contributions, thereby increasing the risk of privacy leakage. To quantify this risk, we formalize the adversarial model as follows.

We consider an \emph{honest-but-curious} adversary at the BS, which follows the prescribed protocol while attempting to infer individual private parameters from the received OTA signals in \eqref{equ:receive signal}. To this end, the adversary proceeds in three steps: first isolating a targeted prosumer $i$’s contribution from the OTA superposition, then recovering the bid $b_i$ from the isolated signal via gain normalization, and finally mapping the recovered bid to the underlying private parameter. Detailed procedures are provided below.

\subsubsection{Signal Isolation}
Exploiting the spatial diversity of the MIMO array, the adversary can design a combining vector for each prosumer to isolate its contribution from $\mathbf{y}^k$. For each prosumer $i$, the adversary applies a unit-norm combining vector $\mathbf{f}_i\in\mathbb{C}^{N_r\times1}$ ($\|\mathbf{f}_i\|_2=1$) to the received signal, yielding
\begin{equation}
\label{equ:r}
    r_{i}^k = \mathbf{f}_i^{\mathsf{H}}\mathbf{y}^k=\sum_{i'\in\mathcal{I}} \mathbf{f}_i^{\mathsf{H}}\mathbf{h}_{i'}x_{i'}^k+\mathbf{f}_i^{\mathsf{H}}\mathbf{z}^k.
\end{equation}
Decomposing \eqref{equ:r} into target, interference, and noise components gives:
\begin{equation}
    r_i^k = \mathbf{f}_i^{\mathsf{H}}\mathbf{h}_ix_i^k + \sum_{i'\neq i}\mathbf{f}_i^{\mathsf{H}}\mathbf{h}_{i'}x_{i'}^k + \mathbf{f}_i^{\mathsf{H}}\mathbf{z}^k.
\end{equation}
Here, the first term is the target component (prosumer $i$'s contribution along the combining direction); the second is the aggregate multiuser interference from all non-target prosumers; and the third term is the projected channel noise. To quantify signal separation, we define the instantaneous power ratio $\gamma_i^k$ of the target signal to the aggregate interference-plus-noise at the combiner output. Applying the Cauchy--Schwarz inequality to upper-bound the interference power in the denominator yields:
\begin{equation}
\begin{aligned}
    \gamma_i^k  &\triangleq\frac{\lvert\mathbf{f}_i^{\mathsf{H}}\mathbf{h}_ix_i^k\rvert^2}{\big\lvert\sum_{i'\neq i}\mathbf{f}_i^{\mathsf{H}}\mathbf{h}_{i'}x_{i'}^k\big\rvert^2+\mathbb{E}[\lvert\mathbf{f}_i^{\mathsf{H}}\mathbf{z}^k\rvert^2 ]} \\
    & \geq \frac{1}{I-1}\cdot\underbrace{\frac{\lvert\mathbf{f}_i^\mathsf{H}{}\mathbf{h}_i\rvert^2\lvert s_{i,1}\rvert^2}{\sum_{i'\neq i}\lvert\mathbf{f}_i^{\mathsf{H}}\mathbf{h}_{i'}\rvert^2\lvert s_{i',1}\rvert^2+\frac{\sigma_z^2}{I-1}}}_{\displaystyle\triangleq\;\overline{\text{SINR}}_i},
\end{aligned}
\end{equation}
where $\overline{\text{SINR}}_i$ is the equivalent signal-to-interference-plus-noise ratio (SINR) for prosumer $i$, which is time-invariant since $\{s_{i,1}\}_{i\in\mathcal{I}}$ and $\mathbf{H}$ are fixed. Since a higher $\overline{\text{SINR}}_i$ implies better signal isolation and thus greater information leakage of prosumer $i$, the adversary optimizes $\mathbf{f}_i$ to maximize $\overline{\text{SINR}}_i$:
\begin{equation}
    \mathbf{f}_i^{\text{adv}} = \arg\max_{\|\mathbf{f}_i\|_2=1}\overline{\text{SINR}}_i(\mathbf{f}_i).
\end{equation}

It has been shown in \cite{kay1993fundamentals} that the minimum mean-square-error (MMSE) estimator maximizes the SINR of such problems. Accordingly, the optimal extractor can be expressed in closed form as 
\begin{equation}
\label{equ:f-adversary}
    \mathbf{f}_i^{\text{adv}} = \frac{\mathbf{B}_i^{-1}\mathbf{h}_i}{\|\mathbf{B}_i^{-1}\mathbf{h}_i\|_2}, 
\end{equation}
where
\begin{equation}
    \mathbf{B}_i=\sum_{i'\neq i}\lvert s_{i',1}\rvert^2\mathbf{h}_{i'}\mathbf{h}_{i'}^{\mathsf{H}}+\frac{\sigma_z^2}{I-1}\mathbf{I}
\end{equation}
denotes the equivalent interference-plus-noise covariance matrix associated with $\overline{\text{SINR}}_i$. The operator $\mathbf{B}_i^{-1}$ whitens the interference-plus-noise space, and the subsequent projection onto $\mathbf{h}_i$ yields the unit-norm direction that maximizes $\overline{\text{SINR}}_i$.


\subsubsection{Bid Recovery via Gain Normalization}
With the optimal extractor $\mathbf{f}_i^{\text{adv}}$, the interference is suppressed and the post-combiner output approximately reduces to a single-prosumer observation:
\begin{equation}
    r_i^k \approx (\mathbf{f}_i^{\text{adv}})^{\mathsf{H}}\mathbf{h}_ix_i^k = \frac{(\mathbf{f}_i^{\text{adv}})^{\mathsf{H}}\mathbf{h}_is_{i,1}}{L}b_i^k.
\end{equation}
Let $g_i\triangleq\frac{(\mathbf{f}_i^{\text{adv}})^{\mathsf{H}}\mathbf{h}_is_{i,1}}{L}$ denote the effective combining gain. The adversary then recovers the bid via gain normalization:
\begin{equation}
    \hat{b}_i^k = \frac{r_i^k}{g_i}.
\end{equation}
\subsubsection{Private-Parameter Inference}
Once the estimated bid $\hat{b}_i^k$ is obtained, the adversary can further infer the prosumer's private parameter $q_i = d_i - p_i$ based on the market-clearing relation in \eqref{equ:3b}. Since the attack is performed at the BS, the global clearing price $\lambda^k$ is known. Therefore, the adversary can estimate the private parameter by
\begin{equation}
    \hat{q}_i^k = -a\lambda^k + \hat{b}_i^k.
\end{equation}
The inference accuracy depends on the SINR achieved by the adversary's optimal extractor, which is governed by the channel noise level and the number of receive antennas $N_r$. Under favorable channel conditions, i.e., when the noise level is low and the antenna array is large, the adversary can recover individual bids and infer prosumers' private parameters with high accuracy, necessitating additional privacy protection. Meanwhile, since the channel noise already partially masks individual signals, the privacy mechanism can be designed to exploit this effect, reducing the artificial perturbation that each prosumer must inject. To this end, we develop a differentially private equilibrium-seeking algorithm in Section~\ref{sec-IV}.

\section{Equilibrium Seeking with Differential Privacy}
\label{sec-IV}
To protect the prosumers' private parameters during the uplink transmission, we develop a differentially private equilibrium seeking algorithm. The proposed approach introduces calibrated random perturbations into the information exchange process, ensuring that the energy sharing mechanism satisfies $(\epsilon,\delta)$-differential privacy. Furthermore, we prove that the algorithm converges in expectation to the generalized Nash equilibrium of the game $\mathcal{G}=(\mathcal{I},\mathcal{S},U)$.

\subsection{Preliminaries on DP}
For each prosumer $i$, let $\mathcal{D}_i$ denote its private local dataset, which includes the parameters of the cost function $f_i(\cdot)$ and utility function $u_i(\cdot)$ that characterize its type and determine the bid $b_i$. Two datasets $\mathcal{D}_i$ and $\mathcal{D}'_i$ are called \textit{adjacent} if they differ in at most one data. In round $k$, prosumer $i$ computes a bid. According to our adversarial model, at the platform, an adversary could extract information $r_i^k$ about prosumer $i$ from received signal vector. Denote the collection of $r_i^k$ across $K$ rounds by
\begin{equation}
    \mathcal{R}_i \triangleq \{r_i^k:1\leq k\leq K\}.
\end{equation}
\begin{definition}[Per-prosumer $(\epsilon_i,\delta_i)$-DP]
The uploading pipeline of prosumer $i$ satisfies $(\epsilon_i,\delta_i)$-DP if, for any pair of adjacent datasets $\mathcal{D}_i$ and $\mathcal{D}'_i$, and any measurable set $\mathcal{O}$:

\begin{equation}
    \Pr(\mathcal{R}_i\in\mathcal{O}|\mathcal{D}_i)\leq e^{\epsilon_i}\Pr(\mathcal{R}_i\in\mathcal{O}|\mathcal{D}_i')+\delta_i
\end{equation}
\end{definition}

Equivalently, the accumulated log-likelihood ratio over $K$ rounds is concentrated:
\begin{equation}
    \Pr\Bigl(\Bigl|\sum_{k=1}^K\ln\frac{\text{Pr}(r_{i}^k|\mathcal{D}_i)}{\text{Pr}(r_{i}^k|\mathcal{D}'_i)}\Bigr|\leq\epsilon_i\Bigr)\geq1-\delta_i
\end{equation}

To protect privacy, each prosumer perturbs its transmitted bid symbol by adding a calibrated random noise to the transmit signal in \eqref{equ:transmit}:
\begin{equation}
\label{equ:DP transmit}
    \tilde{x}_i^k = s_{i,1}\frac{b_i^k}{L} + s_{i,2}n_i^k,\forall i,k,
\end{equation}
where $n_i^k\sim\mathcal{CN}(0,1)$ denotes a normalized complex Gaussian noise. The coefficient $s_{i,2}\in\mathbb{C}$ is calibrated to meet the $(\epsilon_i,\delta_i)$-DP requirement. The power coefficients $s_{i,1}$ and $s_{i,2}$ must satisfy the transmit power constraint
\begin{equation}
\label{equ:power_constraint}
    |s_{i,1}|^2 + |s_{i,2}|^2 \leq P,\quad \forall i,
\end{equation}
this signal is then transmitted to the platform through the OTA channel follows the pipeline we propose in Section \ref{sec:OTA}. 

Similar to \eqref{equ:estimate price}, at the BS, the clearing price estimate under DP mechanism is given by
\begin{equation}
\label{equ:DP price}
    \hat{\lambda}^k_{\text{DP}} = \sum_{i\in\mathcal{I}}\frac{\mathbf{f}_0^{\mathsf{H}}\mathbf{h}_is_{i,1}}{aI\sqrt{\eta}L}b_i^k + \sum_{i\in\mathcal{I}}\frac{\mathbf{f}_0^{\mathsf{H}}\mathbf{h}_is_{i,2}}{aI\sqrt{\eta}}n_i^k + \frac{\mathbf{f}_0^{\mathsf{H}}\mathbf{z}^k}{aI\sqrt{\eta}},
\end{equation}
where the first term represents the useful aggregated bids transmitted by all participating prosumers through the OTA channel, scaled by combining and normalization factors. The second term accounts for the privacy noise independently added by each prosumer before transmission to enforce the DP guarantee, which is then aggregated by the channel and the BS combiner. The third term reflects receiver-side channel noise at the BS.


\subsection{Algorithm Design}
Building on the DP mechanism described above, we now present the complete equilibrium seeking algorithm. Algorithm \ref{alg:alg1} integrates the iterative bidding process with the privacy-preserving OTA communication pipeline. In each iteration, prosumers solve their local optimization problems, encode and perturb their bids using \eqref{equ:DP transmit}, and transmit over the wireless channel. The BS then combines and normalizes the received signals via \eqref{equ:DP price} to update the clearing price estimate $\hat{\lambda}_{\text{DP}}$. 

\begin{algorithm}[!thb]
\SetAlgoVlined
\caption{Energy Sharing Bidding}\label{alg:alg1}
\KwIn{input parameters $f_i(.)$, $u_i(.)$ into each smart meter $i$, tolerance $\nu$.}
\KwOut{energy sharing results $p^{k+1}$, $d^{k+1}$, $b^{k+1}$, $\lambda^{k+1}$.}
\KwInit{$\lambda^1=0$, $k=0$\;}
\Repeat{$|\lambda^{k+1}-\lambda^k| \le \nu$}{
    iteration $k\gets k+1$\;
    \textbf{prosumer update:}\;
    \For{$i\in\mathcal{I}$}{
        $(p_i^{k+1}, d_i^{k+1})$ solves problem \eqref{equ:local} with $\lambda=\lambda^k$\;
        $b_i^{k+1} := d_i^{k+1} - p_i^{k+1} + a\lambda^k$\;
        Encode and perturb bid via \eqref{equ:DP transmit}\;
        Transmit $\tilde{x}_i^{k+1}$ over OTA channel\;
    }
    \textbf{platform update:}\;
    Combine and normalize signals via \eqref{equ:DP price}: $\lambda^{k+1} := \hat{\lambda}^{k+1}_{\text{DP}}$\;
    Broadcast $\lambda^{k+1}$ to all prosumers\;
}
\label{alg1}
\end{algorithm}

Since the algorithm incorporates privacy-preserving perturbations, it is essential to analyze its convergence behavior and quantify how communication design affects both convergence rate and privacy loss. These are addressed in the following subsections.

\subsection{Privacy Budget Analysis}
\label{subsec:privacy_budget}

This subsection analyzes the privacy budget under realistic channel conditions. While traditional differentially private mechanisms assume perfect communication and rely entirely on artificial noise for privacy, the inherent channel noise in our OTA-based framework provides additional privacy amplification. We characterize this benefit by establishing the relationship between the prosumers' noise power allocation and the achievable privacy level, and determine the minimum artificial noise required to meet a target privacy budget.

According to the Gaussian mechanism result in \cite{wei2021user}, prosumer $i$ achieves $(\epsilon_i, \delta_i)$-DP over $K$ iterations if the total noise power in the received signal satisfies
\begin{equation}
\label{equ:DP_bound}
    \sum_{j\in\mathcal{I}} |\mathbf{f}_0^{\mathsf{H}} \mathbf{h}_j|^2 |s_{j,2}|^2 + \sigma_z^2 \geq \Delta_i^2 \frac{2K\ln(1/\delta_i)}{\epsilon_i^2},\end{equation}
where $\Delta_i$ denotes the maximum sensitivity of the disclosed information at the BS. The total noise power on the left-hand side comprises two components: the \emph{aggregated artificial noise} $\sum_{j\in\mathcal{I}} |\mathbf{f}_0^{\mathsf{H}} \mathbf{h}_j|^2 |s_{j,2}|^2$ injected by prosumers, and the \emph{inherent channel noise} $\sigma_z^2$.

The sensitivity $\Delta_i$ measures the maximum influence of prosumer $i$'s private data on the received signal. Since the bid $b_i$ depends on the private dataset $\mathcal{D}_i$, any change in $\mathcal{D}_i$ induces a change in the received signal, due to \eqref{eq:7}:
\begin{equation}
\label{equ:sensitivity}
\Delta_i = \max_{\mathcal{D}_i,\mathcal{D}_i'} \left\|\mathbf{f}_0^{\mathsf{H}}\mathbf{h}_i\frac{s_{i,1}}{L}(b_i(\mathcal{D}_i) - b_i(\mathcal{D}_i'))\right\|_2 \leq 2|\mathbf{f}_0^{\mathsf{H}}\mathbf{h}_i||s_{i,1}|\end{equation}
where the inequality follows from the upper bound derived in \cite{liu2024differentially}. Substituting \eqref{equ:sensitivity} into \eqref{equ:DP_bound}, achieving $(\epsilon_i, \delta_i)$-DP requires the communication parameters to satisfy
\begin{equation}
\label{equ:privacy_bound}
\epsilon_i^2\geq\frac{8|\mathbf{f}_0^{\mathsf{H}}\mathbf{h}_i|^2|s_{i,1}|^2 K \ln(1/\delta_i)}{\sum_{j\in\mathcal{I}}|\mathbf{f}_0^{\mathsf{H}}\mathbf{h}_j|^2|s_{j,2}|^2 + \sigma_z^2}.
\end{equation}

Dividing both numerator and denominator of \eqref{equ:privacy_bound} by $|\mathbf{f}_0^{\mathsf{H}}\mathbf{h}_i|^2|s_{i,1}|^2$ yields
\begin{equation}
\label{equ:privacy_simplified}
\epsilon_i^2 \geq \frac{8 K \ln(1/\delta_i)}{\underbrace{\frac{|s_{i,2}|^2}{|s_{i,1}|^2}}_{\triangleq\,\alpha_i} + \underbrace{\frac{\sum_{j\neq i}|\mathbf{f}_0^{\mathsf{H}}\mathbf{h}_j|^2|s_{j,2}|^2 + \sigma_z^2}{|\mathbf{f}_0^{\mathsf{H}}\mathbf{h}_i|^2|s_{i,1}|^2}}_{\triangleq\,1/\text{SINR}_i}}.
\end{equation}
Here $\alpha_i$ denotes the noise-to-signal ratio at prosumer $i$, and $\text{SINR}_i$ denotes the signal-to-interference-plus-noise ratio. To simplify the multi-user coupling in \eqref{equ:privacy_simplified}, where each prosumer's privacy constraint depends on others' noise power through $\text{SINR}_i$, we adopt a uniform noise-to-signal ratio $\alpha_i \equiv \alpha$ for all prosumers. Under this protocol, $\text{SINR}_i$ becomes a function of the common parameter $\alpha$, which serves as the single control knob for privacy provisioning.

To evaluate the privacy guarantee under the strongest possible attack, we consider a worst-case receiver $\mathbf{f}_i^\star$ that maximizes $\text{SINR}_i$ (see Appendix~\ref{app:privacy_derivation} for the derivation). Denoting the resulting maximum SINR as $\text{SINR}_i^\star(\alpha)$, the privacy bound \eqref{equ:privacy_simplified} becomes
\begin{equation}
\label{equ:privacy_worst_case}
\epsilon_i^2(\alpha) \geq \frac{8 K \ln(1/\delta_i)}{\alpha + 1/\text{SINR}_i^\star(\alpha)}.
\end{equation}
This bound reveals a clear tradeoff: the denominator comprises two components---$\alpha$, the artificial noise ratio that prosumers can control, and $1/\text{SINR}_i^\star(\alpha)$, the inherent system-level masking arising from channel noise and multi-user interference. Stronger privacy (smaller $\epsilon_i$) is achieved by increasing either component.

\textbf{Monotonicity:} The privacy bound \eqref{equ:privacy_worst_case} is monotonically decreasing in $\alpha$. Increasing $\alpha$ raises both the direct $\alpha$ term and $1/\text{SINR}_i^\star(\alpha)$ in the denominator (see Appendix~\ref{app:privacy_derivation} for details), leading to a smaller $\epsilon_i$. 

\textbf{Calibration:} Given target privacy levels $\{\epsilon_{i,\text{target}}\}_{i\in\mathcal{I}}$, the minimum required noise-to-signal ratio is
\begin{equation}
\label{equ:alpha_min}
\alpha_{\min} \triangleq \inf\{\alpha \geq 0 : \epsilon_i(\alpha) \leq \epsilon_{i,\text{target}}, \forall i \in \mathcal{I}\}.
\end{equation}
Due to the monotonicity established above, $\alpha_{\min}$ can be efficiently computed via bisection search.

\subsection{Convergence Analysis}
Building on \eqref{equ:price} and \eqref{equ:DP price}, we define the DP-based estimation error as the discrepancy between the true price and its estimate, $e^k \triangleq\hat{\lambda}^k_{\text{DP}}-\lambda^k$. Specifically, it is given by:
\begin{equation}
    e^k= \frac{1}{aI}\sum_{i\in\mathcal{I}}(\frac{\mathbf{f}_0^{\mathsf{H}}\mathbf{h}_is_{i,1}}{\sqrt{\eta}L}-1)b_i^k+ \sum_{i\in\mathcal{I}}\frac{\mathbf{f}_0^{\mathsf{H}}\mathbf{h}_is_{i,2}}{aI\sqrt{\eta}}n_i^k+\frac{\mathbf{f}_0^{\mathsf{H}}\mathbf{z}^k}{aI\sqrt{\eta}}.
\end{equation}

We further decompose the error as $e^k=e^k_{\text{align}}+e^k_{\text{noise}}$, with 
\begin{equation}
\label{equ:align}
    e^k_{\text{align}} \triangleq\frac{1}{aI}\sum_{i\in\mathcal{I}}(\frac{\mathbf{f}_0^{\mathsf{H}}\mathbf{h}_is_{i,1}}{\sqrt{\eta}L}-1)b_i^k,
\end{equation}
\begin{equation}
\label{equ:noise}
    e^k_{\text{noise}} \triangleq  \sum_{i\in\mathcal{I}}\frac{\mathbf{f}_0^{\mathsf{H}}\mathbf{h}_is_{i,2}}{aI\sqrt{\eta}}n_i^k+\frac{\mathbf{f}_0^{\mathsf{H}}\mathbf{z}^k}{aI\sqrt{\eta}},
\end{equation}
where the term $e^k_{\text{align}}$ represents the signal mismatch in OTA aggregation, it is design-removable via proper pre-equalization (details provided later). The noise error term $e^k_{\text{noise}}$ collects the injected artificial noise to protect privacy and communication noise after combining and normalization. Unlike the alignment term, it persists even under perfect alignment. 

To characterize the convergence behavior, we first list the standard assumptions on the prosumers' utility and cost functions.

\textbf{Assumption 1.} For each prosumer $i\in\mathcal{I}$, the utility function $u_i(d_i)$ is $\mu_u$-strongly concave and the cost function $f_i(p_i)$ is $\mu_f$-strongly convex for some $\mu_u,\mu_f>0$.

\textbf{Assumption 2.} The gradients $\nabla u_i(\cdot)$ and $\nabla f_i(\cdot)$ are Lipschitz continuous with constants $L_u$ and $L_f$, respectively.

Under these standard assumptions on convex games, we derive the following convergence result for Algorithm \ref{alg:alg1}.

\begin{theorem}\label{theorem-1}
Suppose Assumptions 1--2 hold. Let $\lambda^\star$ denote the equilibrium price of the energy sharing game $\mathcal{G}=(\mathcal{I},\mathcal{S},U)$. If the market sensitivity parameter satisfies
\begin{equation}
\label{equ:a_condition}
a > \frac{I-3}{2(I-1)}\left(\frac{1}{\mu_u}+\frac{1}{\mu_f}\right),
\end{equation}
then under Algorithm~\ref{alg:alg1}, the expected squared distance to equilibrium satisfies
\begin{equation}
\label{equ:convergence}
\mathbb{E}[|\lambda^K - \lambda^\star|^2] \leq \rho^K |\lambda^0 - \lambda^\star|^2 + \frac{1-\rho^K}{1-\rho}\mathbb{E}[|e^k_{\text{noise}}|^2],
\end{equation}
where $\rho = \left(1 - \frac{(I-1)\gamma}{a(I-1)+\gamma}\right)^2 \in (0,1)$ is the contraction factor with $\gamma \in \left[\frac{1}{L_u}+\frac{1}{L_f},\; \frac{1}{\mu_u}+\frac{1}{\mu_f}\right]$. Specifically, when $f_i(\cdot)$ and $u_i(\cdot)$ are quadratic for all $i\in\mathcal{I}$, the algorithm is asymptotically unbiased in mean, i.e., $\lim_{K\to\infty}\mathbb{E}[\lambda^K]=\lambda^\star$.
\end{theorem}

The proof of Theorem \ref{theorem-1}, including its quadratic special case, is given in  Appendix~\ref{app:proof_theorem1}. Theorem \ref{theorem-1} reveals that the convergence performance is fundamentally limited by the DP-induced noise. The first term on the right-hand side represents the residue from initialization, which decreases exponentially with the iteration number $K$. The second term indicates a non-diminishing loss directly determined by $e^k_{\text{noise}}$ in \eqref{equ:noise}, which comprises the artificial noise for privacy protection and the communication noise. Therefore, the convergence rate is inherently affected by the privacy-preserving mechanism, and faster convergence requires reducing the noise term $e^k_{\text{noise}}$.

\section{Case Study}
\label{sec-V}
In this section, we present numerical results to evaluate the proposed OTA-based differentially private energy sharing mechanism. We first describe the simulation setup, then evaluate the privacy protection effectiveness and the benefits of mechanism-communication co-design. Subsequently, we analyze the convergence behavior and compare OTA aggregation with orthogonal multiple-access schemes. Finally, we validate the algorithm under realistic 3GPP channel models.

\subsection{Simulation Setup}
We consider a community of three prosumers participating in the iterative energy sharing mechanism described in this paper. For each prosumer $i\in\mathcal{I}=\{1,2,3\}$, the production cost and demand utility are quadratic and given by $f_i(p_i)=c_{1,i}p_i^2+c_{2,i}p_i$ and $u_i(d_i)=v_{1,i}d_i^2+v_{2,i}d_i$, respectively, where the coefficients $\{c_{1,i}, c_{2,i},v_{1,i},v_{2,i}\}_{i\in\mathcal{I}}$ are summarized in Table~\ref{tab:coef}, the market sensitivity parameter is fixed at $a = 100$. For the wireless communication setup, we adopt a Rayleigh fading channel model, where the channel coefficient vectors $\mathbf{h}_i$ are drawn from $\mathcal{CN}(\mathbf{0}, \mathbf{I}_{N_r})$. The transmit power budget for each prosumer is $P = 1$ W. The channel noise variance $\sigma_z^2$ is determined by the signal-to-noise ratio (SNR), defined as $\text{SNR} = P / \sigma_z^2$ or equivalently $\text{SNR}\big|_{\text{dB}} = 10\log_{10}(P/\sigma_z^2)$. For differential privacy, we fix $\delta = 10^{-5}$ throughout all experiments. All results are averaged over 100 Monte Carlo trials with independent channel realizations.



\begin{table}[t]
  \centering
  \caption{Cost coefficients of prosumers.}
  \label{tab:coef}
  \begin{tabular}{ccccc}
    \toprule
    \multirow[b]{2}{*}{Prosumer}
      & $c_{1,i}$
      & $c_{2,i}$
      & $v_{1,i}$
      & $v_{2,i}$ \\
      & ($\$/\mathrm{kWh}^2$)
      & ($\$/\mathrm{kWh}$)
      & ($\$/\mathrm{kWh}^2$)
      & ($\$/\mathrm{kWh}$) \\
    \midrule
    1 & 0.018 & 0.025 & $-0.006$ & 0.9 \\
    2 & 0.012 & 0.065 & $-0.008$ & 0.7 \\
    3 & 0.014 & 0.045 & $-0.007$ & 0.6 \\
    \bottomrule
  \end{tabular}
\end{table}

\subsection{Privacy Protection Evaluation}
\begin{figure*}[!t]
\centering
\includegraphics[width=0.95\textwidth]{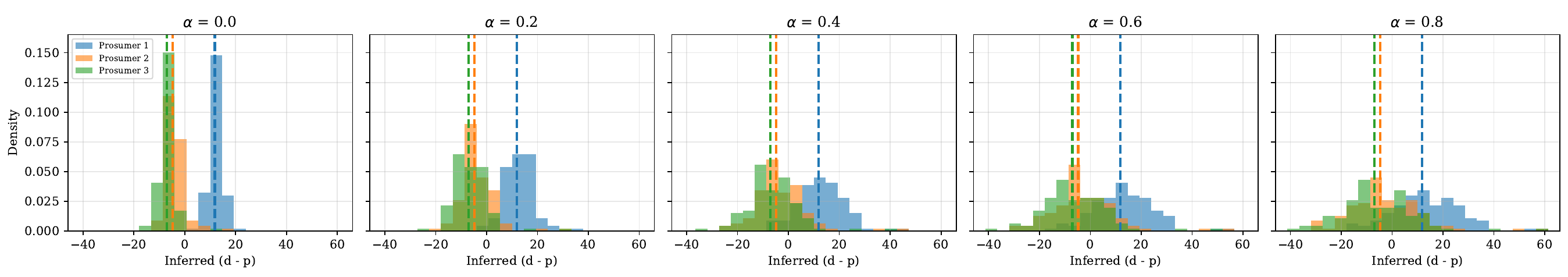}
\caption{Distribution of adversary's inferred private parameter $(d_i - p_i)$ for each prosumer under different noise-to-signal ratios ($\alpha = 0, 0.2, 0.4, 0.6, 0.8$). The dashed vertical lines indicate the true values. Higher $\alpha$ leads to more dispersed distributions, demonstrating the effectiveness of the DP mechanism.}
\label{fig:attack}
\end{figure*}

To evaluate the effectiveness of the proposed differential privacy mechanism against adversarial attacks, we simulate the attack model described in Section~\ref{subsec:attack_model} under different noise-to-signal ratios $\alpha$. Fig.~\ref{fig:attack} shows the distribution of the adversary's inferred net demand $d_i - p_i$ for each prosumer across multiple attack trials.

As the noise-to-signal ratio $\alpha$ increases from $0$ to $0.8$, the distribution of inferred values becomes increasingly dispersed around the true value, and the attack success rate decreases significantly. This demonstrates that the calibrated artificial noise effectively protects prosumers' private information by degrading the adversary's inference accuracy.

\begin{figure*}[!t]
\centering
\includegraphics[width=0.95\textwidth]{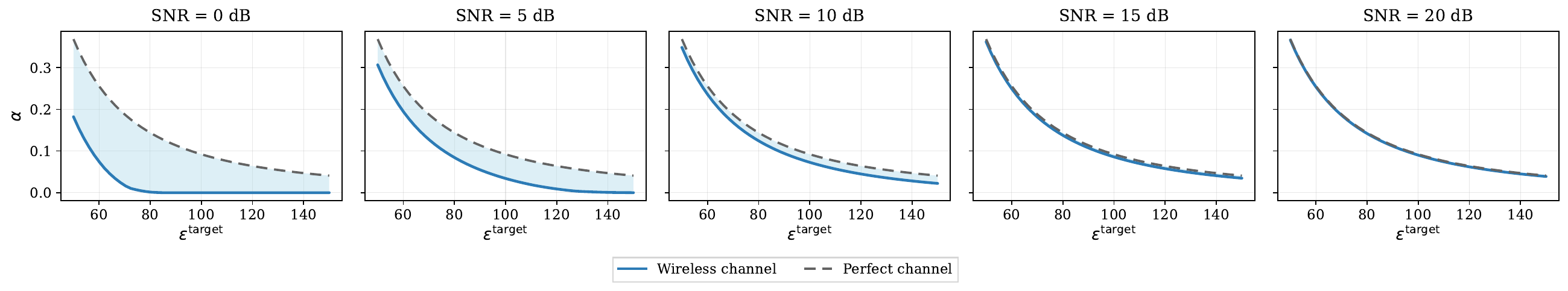}\\[2pt]
\includegraphics[width=0.95\textwidth]{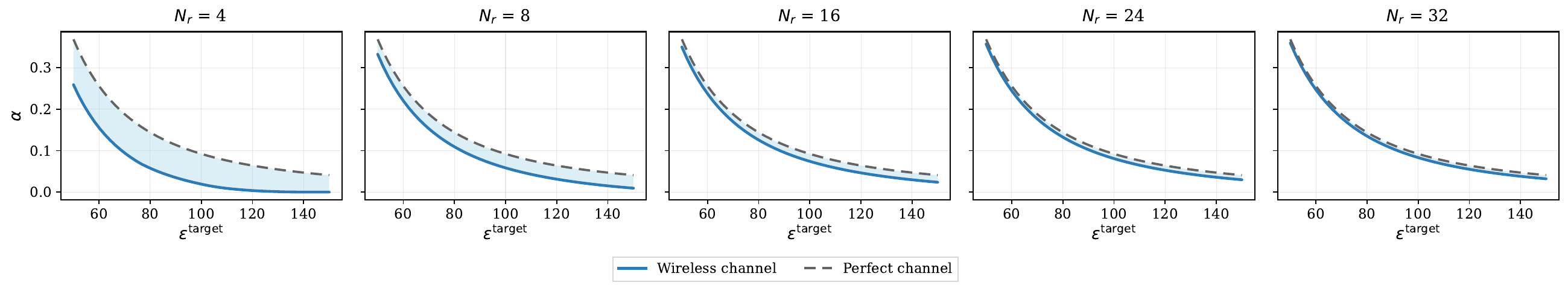}
\caption{Artificial noise-to-signal ratio $\alpha$ versus target privacy budget $\epsilon^{\text{target}}$, comparing a perfect channel (no channel noise) and wireless channels. The shaded area represents the artificial noise saving enabled by channel noise. Top: different SNR values with fixed $N_r = 8$. Bottom: different $N_r$ values with fixed SNR $= 10$ dB.}
\label{fig:alpha_scan}
\end{figure*}

While the DP mechanism is effective, a key question is how much artificial noise is actually needed. The answer depends on the communication channel: by jointly designing the market mechanism with the wireless transmission, we can exploit the inherent channel noise to reduce the required artificial noise level. To demonstrate this co-design advantage, we compare the minimum artificial noise-to-signal ratio $\alpha$ required to achieve a target privacy budget $\epsilon^{\text{target}}$ under wireless channels versus a perfect channel (no channel noise). The results, shown in Fig.~\ref{fig:alpha_scan}, reveal that exploiting channel impairments can significantly reduce the required artificial noise level. Across all configurations, the wireless channel requires a smaller $\alpha$ than the perfect channel to achieve the same privacy level, with the gap (shaded area) representing the artificial noise saving. This saving is influenced by two key factors: (i) SNR (Fig.~\ref{fig:alpha_scan}, top): as SNR decreases, channel noise provides stronger privacy protection, reducing $\alpha$ by more than half at SNR $= 0$ dB (i.e., the noise power equals the transmit power); (ii) number of antennas $N_r$ (Fig.~\ref{fig:alpha_scan}, bottom): as $N_r$ increases, the adversary's signal separation capability improves, reducing the privacy benefit from channel noise. These results validate the theoretical analysis in Section~\ref{subsec:privacy_budget} and demonstrate the practical benefits of mechanism-communication co-design.

\subsection{Convergence Analysis}
\begin{figure*}[!t]
\centering
\includegraphics[width=\textwidth]{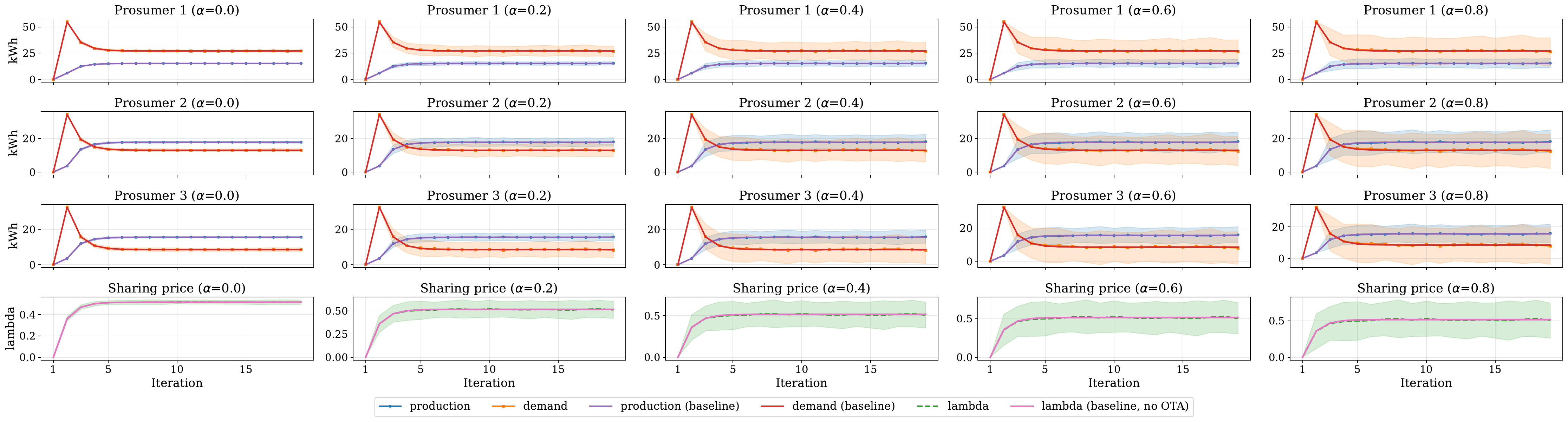}
\caption{Convergence of production, demand, and clearing price under OTA aggregation with different noise-to-signal ratios ($\alpha = 0, 0.2, 0.4, 0.6, 0.8$). Shaded regions indicate 95\% CI over 100 trials.}
\label{fig:convergence}
\end{figure*}

To validate the convergence guarantees established in Theorem~1, we evaluate the algorithm under different artificial noise-to-signal ratios $\alpha$ ranging from 0 to 0.8. Fig.~\ref{fig:convergence} shows the evolution of production $p_i$, demand $d_i$, and clearing price $\lambda$ over iterations.

The results demonstrate two key observations. First, the algorithm converges to the generalized Nash equilibrium in expectation across all values of $\alpha$, consistent with the asymptotic unbiasedness established in Theorem~\ref{theorem-1} under the quadratic model. The mean trajectories (solid lines) converge to the same equilibrium values regardless of $\alpha$. Second, the convergence variance increases with $\alpha$, as indicated by the widening shaded regions (95\% confidence intervals). This is consistent with Theorem~1, which shows that the mean-square deviation is bounded proportionally to the noise power $\mathbb{E}[|e^k_{\text{noise}}|^2]$.

\begin{figure}[t]
\centering
\includegraphics[width=0.55\linewidth]{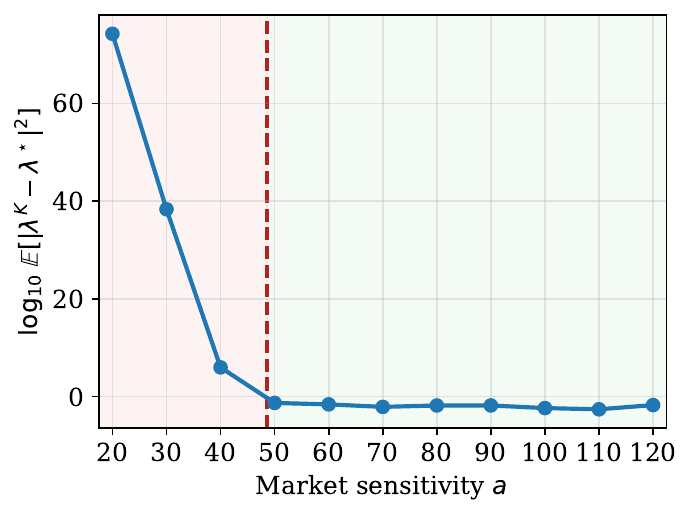}
\caption{Mean squared error $\mathbb{E}[|\lambda^K - \lambda^*|^2]$ (log-scale) versus market sensitivity $a$ with $I=10$ prosumers after $K = 100$ iterations with $\alpha = 0.2$. The dashed line marks the exact convergence boundary where $\rho = 1$. The shaded red (left) and green (right) regions indicate divergence and convergence, respectively.}
\label{fig:a_scan}
\end{figure}

Fig.~\ref{fig:a_scan} validates the convergence condition in Theorem~\ref{theorem-1} using $I=10$ prosumers by plotting $\log_{10}\mathbb{E}[|\lambda^K - \lambda^*|^2]$ after $K=100$ iterations versus $a$ with $\alpha = 0.2$. Each point is averaged over 100 Monte Carlo trials. For small $a$, the algorithm fails to converge; as $a$ increases, the MSE decreases and reaches near-zero levels, confirming that a sufficiently large $a$ ensures $\rho \in (0,1)$ and thus convergence, as predicted by Theorem~\ref{theorem-1}.

\subsection{Privacy Comparison of Communication Schemes}
OTA aggregation provides an additional privacy advantage over conventional orthogonal communication schemes (e.g., TDMA/FDMA) through multi-user interference. In orthogonal transmission, each prosumer transmits in a dedicated time or frequency slot, and the platform receives individual signals without inter-user interference. In contrast, OTA aggregation superposes all prosumers' signals over the air, creating mutual interference that masks individual information. Table~\ref{tab:ota_vs_ortho} compares the privacy budget $\varepsilon$ under various communication configurations, where $\varepsilon_{\text{Ortho}}$ is determined by artificial noise alone, while $\varepsilon_{\text{OTA}}$ is evaluated under the attack model in Section~\ref{subsec:attack_model}.

The results show that OTA consistently achieves equal or better privacy (lower $\varepsilon$) than orthogonal transmission. The privacy gain increases with the loading ratio $I/N_r$, reaching Gain $= 1.24$ in the overloaded regime, as the adversary's spatial resolution degrades. Conversely, at high SNR or when $N_r \gg I$, signal separation becomes easy and OTA's advantage vanishes. These results confirm that OTA achieves privacy amplification via multi-user interference, with the magnitude determined by the loading ratio and channel conditions.

\definecolor{colOrtho}{RGB}{240,240,240}
\definecolor{colOTA}{RGB}{225,225,225}
\begin{table}[t]
  \centering
  \caption{Privacy Budget Comparison: Orthogonal Transmission vs. OTA Aggregation}
  \label{tab:ota_vs_ortho}
  \renewcommand{\arraystretch}{1.05}
  \begin{tabular}{@{}c ccc >{\columncolor{colOrtho}}c >{\columncolor{colOTA}}c c@{}}
  \toprule
  \multirow{2}{*}[-4pt]{Scenario} & \multicolumn{3}{c}{System Parameters} & \multicolumn{3}{c}{Privacy Budget ($\varepsilon$)} \\
  \cmidrule(lr){2-4} \cmidrule(lr){5-7}
   & $I$ & $N_r$ & SNR (dB) & Ortho. & OTA & Gain \\
  \midrule
  Baseline      & 3  & 8  & 10 & 97.63  & 94.31  & 1.04 \\
  Low SNR       & 3  & 8  & 0  & 46.22  & 45.53  & 1.02 \\
  High SNR      & 3  & 8  & 30 & 124.07 & 123.91 & 1.00 \\
  Crowded       & 8  & 8  & 10 & 57.43  & 50.37  & 1.14 \\
  Overloaded    & 12 & 8  & 10 & 50.89  & 41.06  & 1.24 \\
  Many Antennas & 3  & 64 & 10 & 135.46 & 135.32 & 1.00 \\
  \bottomrule
  \end{tabular}

  \small
  \textit{Note:} Gain $\triangleq \varepsilon_{\text{Ortho}} / \varepsilon_{\text{OTA}}$. Values $>1$ indicate OTA achieves stronger privacy. Fixed: $\alpha = 0.2$, $\delta = 10^{-5}$.
\end{table}

\subsection{Evaluation under Realistic Channels}
\begin{figure}[t]
\centering
\begin{minipage}[b]{0.48\linewidth}
\centering
\includegraphics[height=3.7cm]{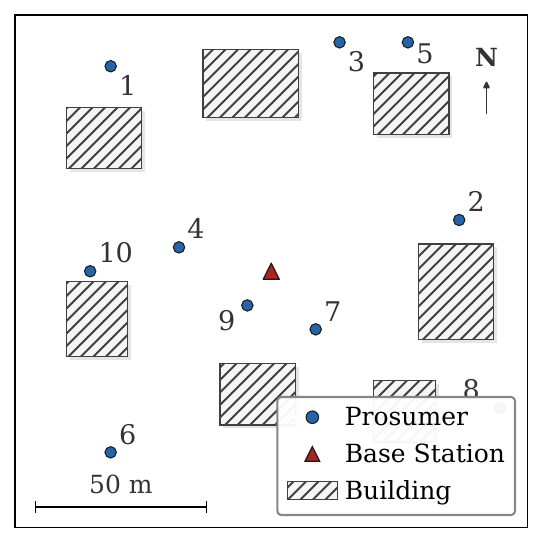}
\end{minipage}\hspace{-2mm}
\begin{minipage}[b]{0.48\linewidth}
\centering
\includegraphics[height=3.7cm]{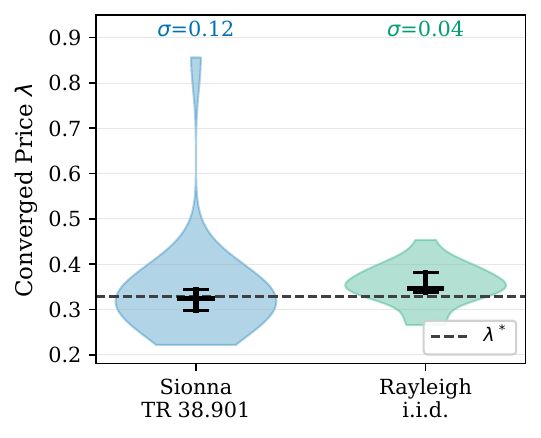}
\end{minipage}
\caption{Evaluation under realistic channels. Left: spatial distribution of prosumers in the Sionna simulation. Right: violin plot comparing the converged price $\lambda$ distribution under Sionna TR 38.901 and i.i.d. Rayleigh channel models; the dashed line indicates $\lambda^*$ and the black bars represent the interquartile range with the median.}
\label{fig:sionna}
\end{figure}

The previous experiments assume i.i.d. Rayleigh fading channels. To validate the robustness of our approach under more realistic conditions, we replace the Rayleigh channel model with a ray-tracing based channel generated by Sionna~\cite{sionna}. Specifically, we adopt the 3GPP TR 38.901 5G NR channel model with the Urban Microcell (UMi) Street Canyon scenario, operating at 3.5 GHz (Sub-6 GHz) carrier frequency for uplink transmission. As shown in Fig.~\ref{fig:sionna} (left), we consider a customized urban area where prosumers are distributed at different locations. Due to the presence of buildings and obstacles, some prosumers experience severe signal blockage, resulting in significantly degraded channel conditions compared to others.

Fig.~\ref{fig:sionna} (right) compares the converged price distribution under the two channel models. The Sionna model exhibits a notably larger variance and a longer tail compared to the i.i.d. Rayleigh model, due to spatial correlation in realistic urban propagation. Despite this, both models converge to prices near the Nash equilibrium $\lambda^*$ on average, validating the robustness of the proposed algorithm under realistic channel conditions.

\section{Conclusion}
\label{sec-VI}
In this paper, we develop a mechanism and communication co-design framework for differentially private energy sharing, which jointly addresses game-theoretic market design and OTA MIMO communication. We formulate the energy sharing problem as a generalized Nash game and show that implementing the iterative bidding process over wireless OTA MIMO channels exposes prosumers to non-trivial privacy risks, as quantified by an honest-but-curious adversarial model at the base station. To mitigate such leakage, we propose a differentially private equilibrium-seeking algorithm that superposes calibrated artificial noise on prosumer bids before OTA aggregation. We prove that the resulting iterative process converges in expectation to a neighborhood of the equilibrium.

Simulations demonstrate the effectiveness of the proposed scheme with the following findings: (1) By exploiting inherent channel noise, the co-design approach reduces the required artificial perturbation by as much as half compared to methods assuming ideal communication, lowering the privacy protection overhead for prosumers. (2) OTA aggregation achieves up to 24\% stronger privacy protection than conventional orthogonal schemes such as TDMA/FDMA, with the advantage growing as more prosumers participate in energy sharing. (3) The mean-square deviation of the algorithm from the generalized Nash equilibrium remains bounded under all tested privacy levels. Specifically, under the quadratic model, the algorithm is asymptotically unbiased in mean, and the mean trajectories converge to the equilibrium, preserving near-optimal energy sharing performance.

Future work includes extending the framework to dynamic channel conditions with imperfect channel state information, and investigating more sophisticated adversarial models such as colluding prosumers or active attackers.

\bibliographystyle{IEEEtran}
\bibliography{refs}

\appendices
\setcounter{equation}{0}
\renewcommand{\theequation}{A.\arabic{equation}}

\section{Derivation of Optimal Information Extractor}
\label{app:privacy_derivation}

This appendix provides the detailed derivation of the optimal information extractor in Section~\ref{subsec:privacy_budget}.

\textbf{Step 1: Simplification to SINR maximization.} For a given power allocation, the optimal extractor maximizes the privacy bound \eqref{equ:privacy_bound}:
\begin{align}
\mathbf{f}_i^\star &= \arg\max_{\|\mathbf{f}_i\|_2=1} \frac{8|\mathbf{f}_i^{\mathsf{H}}\mathbf{h}_i|^2|s_{i,1}|^2 K \ln(1/\delta_i)}{\sum_{j\in\mathcal{I}}|\mathbf{f}_i^{\mathsf{H}}\mathbf{h}_j|^2|s_{j,2}|^2 + \sigma_z^2} \notag \\
&= \arg\max_{\|\mathbf{f}_i\|_2=1} \frac{|\mathbf{f}_i^{\mathsf{H}} (\mathbf{h}_i s_{i,1})|^2}{\sum_{j\neq i}|\mathbf{f}_i^{\mathsf{H}} (\mathbf{h}_j s_{j,2})|^2 + \sigma_z^2},
\end{align}
where we remove the constant terms independent of $\mathbf{f}_i$. The final form is a standard SINR maximization problem.

\textbf{Step 2: Closed-form solution.} Under the uniform noise-to-signal ratio $\alpha_i \equiv \alpha$ adopted in Section~\ref{subsec:privacy_budget}, we have $|s_{j,2}|^2 = \alpha |s_{j,1}|^2$ for all $j$. Note that $|s_{j,1}|^2$ is not an independent parameter: under perfect OTA pre-equalization and the maximal feasible normalization under the power constraint \eqref{equ:power_constraint}, it is uniquely determined by $\alpha$ and the fixed channel realization and combiner. Therefore, $\alpha$ is the only remaining tunable parameter, and all quantities below are implicit functions of $\alpha$ alone. The interference-plus-noise covariance matrix becomes
\begin{equation}
\mathbf{B}_i(\alpha) = \alpha \sum_{j\neq i}|s_{j,1}|^2\mathbf{h}_j\mathbf{h}_j^{\mathsf{H}} + \sigma_z^2\mathbf{I}.
\end{equation}
By the MMSE criterion \cite{kay1993fundamentals}, the optimal extractor is
\begin{equation}
\mathbf{f}_i^\star(\alpha) = \frac{\mathbf{B}_i^{-1}(\alpha)\mathbf{h}_i}{\|\mathbf{B}_i^{-1}(\alpha)\mathbf{h}_i\|_2}.
\end{equation}

\textbf{Step 3: Maximum SINR.} Substituting $\mathbf{f}_i^\star(\alpha)$ into the SINR expression yields
\begin{equation}
\text{SINR}_i^\star(\alpha) = |s_{i,1}|^2\mathbf{h}_i^{\mathsf{H}}\mathbf{B}_i^{-1}(\alpha)\mathbf{h}_i.
\end{equation}

\textbf{Step 4: Monotonicity in $\alpha$.} Writing $\mathbf{B}_i(\alpha) = \alpha \mathbf{A}_i + \sigma_z^2\mathbf{I}$ where $\mathbf{A}_i = \sum_{j\neq i}|s_{j,1}|^2\mathbf{h}_j\mathbf{h}_j^{\mathsf{H}} \succeq 0$, we observe that increasing $\alpha$ increases $\mathbf{B}_i(\alpha)$ in the positive semidefinite order. Consequently, $\mathbf{B}_i^{-1}(\alpha)$ decreases in the positive semidefinite order, and hence $\text{SINR}_i^\star(\alpha)$ is monotonically decreasing in $\alpha$. Combined with the direct $\alpha$ term in the denominator of \eqref{equ:privacy_worst_case}, this establishes the monotonicity of the privacy bound claimed in Section~\ref{subsec:privacy_budget}.

\textbf{Step 5: Privacy bound.} Substituting $\text{SINR}_i^\star(\alpha)$ into the simplified privacy condition yields \eqref{equ:privacy_worst_case}.

\setcounter{equation}{0}
\renewcommand{\theequation}{B.\arabic{equation}}
\section{Proof of Theorem~\ref{theorem-1}}
\label{app:proof_theorem1}

We prove Theorem~\ref{theorem-1} by analyzing the price update dynamics under noisy price estimates. From the price update rule~\eqref{equ:price}, we define the price mapping
\begin{equation}
T(\lambda) \triangleq \frac{1}{aI}\sum_{i\in\mathcal{I}} b_i(\lambda),
\end{equation}
where $b_i(\lambda) = d_i(\lambda) - p_i(\lambda) + a\lambda$ and $(p_i(\lambda), d_i(\lambda))$ is the unique maximizer of~\eqref{equ:local} for a given price $\lambda$. The ideal price dynamics follow $\lambda^{k+1} = T(\lambda^k)$, while under noisy estimates:
\begin{equation}
\lambda^{k+1} = T(\lambda^k) + e^{k+1}_{\text{noise}}.
\end{equation}

To establish the contraction property of $T$, we use the first-order optimality conditions of~\eqref{equ:local}: $f_i'(p_i)=u_i'(d_i)= \lambda + \frac{q_i}{a(I-1)}$, where $q_i := d_i - p_i$. By the implicit function theorem:
\begin{equation}
\frac{\mathrm{d} q_i}{\mathrm{d}\lambda} = -\frac{a(I-1)\gamma_i}{a(I-1)+\gamma_i},
\end{equation}
where $\gamma_i := \frac{1}{f_i''(p_i)} - \frac{1}{u_i''(d_i)} \in \left[\frac{1}{L_u}+\frac{1}{L_f},\;\frac{1}{\mu_u}+\frac{1}{\mu_f}\right]$. The derivative of $T$ is
\begin{equation}
T'(\lambda) = 1 - \frac{1}{I}\sum_{i\in\mathcal{I}}\frac{(I-1)\gamma_i}{a(I-1)+\gamma_i}.
\end{equation}
For contraction we need $|T'(\lambda)| < 1$. Since each summand $\frac{(I-1)\gamma_i}{a(I-1)+\gamma_i}$ is positive, $T'(\lambda) < 1$ holds automatically. For $T'(\lambda) > -1$, the monotonicity of $\frac{(I-1)\gamma}{a(I-1)+\gamma}$ in $\gamma$ gives the worst-case requirement
\begin{equation}
\frac{(I-1)\gamma_{\max}}{a(I-1)+\gamma_{\max}} < 2,
\end{equation}
where $\gamma_{\max} = \frac{1}{\mu_u}+\frac{1}{\mu_f}$, which is precisely condition~\eqref{equ:a_condition}. Let $L_T = \sup_\lambda |T'(\lambda)| < 1$ and $\rho \triangleq L_T^2 \in (0,1)$.

Define $\tau^k \triangleq \lambda^k - \lambda^\star$. The noisy dynamics give
\begin{equation}
\tau^{k+1} = T(\lambda^k) - T(\lambda^\star) + e^{k+1}_{\text{noise}}.
\end{equation}
By the mean value theorem, $|T(\lambda^k) - T(\lambda^\star)| \leq L_T |\tau^k|$. Taking squares and expectations, the cross term vanishes by independence and zero mean:
\begin{equation}
\mathbb{E}[|\tau^{k+1}|^2] \leq \rho \mathbb{E}[|\tau^k|^2] + \mathbb{E}[|e^{k+1}_{\text{noise}}|^2].
\end{equation}

Expanding the recursion over $K$ iterations with stationary noise $\sigma_e^2 = \mathbb{E}[|e^k_{\text{noise}}|^2]$:
\begin{equation}
\mathbb{E}[|\lambda^K - \lambda^\star|^2] \leq \rho^K|\lambda^0-\lambda^\star|^2 + \frac{1-\rho^K}{1-\rho}\sigma_e^2,
\end{equation}
which completes the proof of the general MSE bound.

For the quadratic special case, $f_i''$ and $u_i''$ are constants, 
so $\gamma_i$ in (B.3) is constant. By (B.4),
\begin{equation}
T'(\lambda) = 1 - \frac{1}{I}\sum_{i\in\mathcal{I}} 
\frac{(I-1)\gamma_i}{a(I-1)+\gamma_i} \triangleq \kappa
\end{equation}
is a constant, and hence $T$ is affine. Since $T(\lambda^\star)=\lambda^\star$, 
the noisy recursion (B.2) becomes
\begin{equation}
\lambda^{k+1} - \lambda^\star = \kappa(\lambda^k - \lambda^\star) 
+ e_{\mathrm{noise}}^{k+1}.
\end{equation}
Taking expectations and using $\mathbb{E}[e_{\mathrm{noise}}^{k+1}]=0$ yields
\begin{equation}
\mathbb{E}[\lambda^{k+1}] - \lambda^\star 
= \kappa\bigl(\mathbb{E}[\lambda^k] - \lambda^\star\bigr).
\end{equation}
By recursion, $\mathbb{E}[\lambda^K] - \lambda^\star 
= \kappa^K(\lambda^0 - \lambda^\star)$. 
Under condition (38), $|\kappa|<1$, so 
$\lim_{K\to\infty}\mathbb{E}[\lambda^K] = \lambda^\star$. 
This completes the proof. $\square$

\newpage

 




\vfill

\end{document}